\documentstyle[preprint,prl,aps]{revtex}
\begin{document}
\preprint{\vbox{\hbox {September 1996} \hbox{HUTP-96/A044} \hbox{IFP-735-UNC} }}
\title{Cabibbo Mixing and the Search for CP Violation}
\author{\bf Paul H. Frampton
$^{(a)}$
and Sheldon L. Glashow$^{(b)}$}
\address{$^{(a)}$
University of North Carolina, Chapel Hill, NC  27599-3255}
\address{$^{(b)}$ Harvard University, Cambridge, MA 02138.}
\maketitle
\begin{abstract}
We examine certain extensions of the standard model in which $CP$
violation is spontaneous and the strong $CP$ problem is resolved. In these
models, 
the $3\times 3$ quark mixing matrix
is neither real nor unitary. However, to a precision 
of~0.1\%, it is real and orthogonal.
There are no readily observable $CP$-violating effects 
besides those in the neutral kaon system. 
\end{abstract}
\newpage

The Standard Model is often regarded as a correct and complete description
of all elementary-particle phenomena, but its depiction of $CP$ violation
is uncertain. Although the effect was first observed three decades
ago\cite{FC}, only one $CP$-violating parameter ($\epsilon_K$ for neutral
kaons) is precisely measured. Others are indecisively measured ({\it e.g.,}
$\epsilon'/\epsilon$) or merely constrained ({\it e.g.,} the  electric
dipole moment of the neutron).  More data is essential to achieve a
definitive theory of $CP$ violation, such as forthcoming  studies of
neutral $B$ decays that will contribute to  the evaluation of the
parameters of quark mixing and provide important tests of the standard
model \cite{BS1,BS2,NQ}. However, the strong $CP$ problem  cannot be solved
in the context of the standard model. Put in a nutshell, there is no
natural way to control the phase of the determinant of the quark mass
matrix  without imposing  $CP$ conservation on the Lagrangian.

In this paper, we explore $CP$-violation in the context of an `Aspon model'
\cite{FK,FN,FKNW,AFKL}  wherein the
violation of $CP$ is spontaneous. In particular, we add to the 3-family
standard model a vector-like SU(2) doublet of quarks $(U, D)_L \oplus    
(U, D)_R$ together with two complex $SU(2)$ singlet 
scalars $\chi_{1,2}.$\footnote {What we say is applicable to 
an alternative Aspon model where
the added quarks are
$SU(2)$ singlets.} 
 Under an additional  gauged $U(1)'$, these
four representations carry the quantum numbers 
$X= +1,\, -1,\, +1,\, +1$, respectively. All conventional fields carry $X=0$.
 $CP$ invariance of the Lagrangian is assumed. Thus the Yukawa couplings of
$\chi_i$ and of conventional Higg(s) fields are real. 
Spontaneous $CP$ violation arises from complex 
vacuum expectation values acquired by the scalar singlets:
$\langle \chi_1\rangle = \kappa_1$ 
and $\langle\chi_2\rangle = \kappa_2$.
 
In our scheme, quark masses are described
by  $4\times 4$ matrices ${\cal M}_{u,d}$ taking the form:
\begin{equation}
{\cal M}_{u,d} = \left( \begin{array}{cc} m_{u,d} & F \cr 
0 & M \cr \end{array} \right),\label{Mass}
\end{equation}
where $m_{u,d}$ are {\it real\/} $3\times3$ matrices  resulting from the
conventional Higgs mechanism and
 $M$ is the Dirac mass
of the added quark doublet. $CP$ violation results from 
the complex off-diagonal matrix elements
$F$ that form  a column vector. Its components:
\begin{equation}
F_k = h_k^1\kappa_1 +  h_k^2\kappa_2  .
\end{equation}
involve the complex VEV's $\kappa$ of the $\chi$ fields and their
real Yukawa couplings:
\begin{equation}
L_{\rm Yukawa} = \sum_{k=1}^3\left(h_k^1\chi_1 +h_k^2\chi_2\right)\left(
{\bar{u}}_L^kU_R + {\bar{d}}_L^kD_R\right)
+ {\it h.c.}
\end{equation}
We define the parameters $x_k$ (assumed to be small) to be $x_k=F_k/M$. 

The up and down  mass matrices are diagonalized by bi-unitary
transformations:
\begin{equation}
J_L^{\dagger}{\cal M}_u J_R = {\rm diag}(m_u, m_c, m_t, M^{\prime}),
\end{equation}
\begin{equation}
\qquad K_L^{\dagger}{\cal M}_dK_R = {\rm diag}(m_d, m_s, m_b, M^{\prime\prime}),
\label{diag}
\end{equation}
where we hereafter ignore the small differences between $M$, $M^\prime$ and
$M^{\prime\prime}$. The
quark flavor mixing matrix is $C=J_L^\dagger K_L$, or 
more explicitly, $C_{\mu\nu} =J_{L\lambda\mu}^*K_{L\lambda\nu}$, where
indices 1,2,3 (henceforth Latin) refer to known quarks and subscript 4 
refers to the added heavy quarks. 
The $CP$-violating phases in 
$C_{ij}$ are suppressed by the small parameters $x_k^2$ (the subscript
is hereafter dropped). 

A lower bound on $x^2$ arises from the need to describe observed $CP$
violation in the neutral kaon system.
Using Eq.(35) of \cite{FN} and requiring that $U(1)'$
break down above the electroweak scale
({\it i.e.,} $\kappa \geq 250$~GeV), we find $x^2 \geq
3 \times 10^{-5}$. With $x^2$ this large,
box diagrams with $\chi$ exchange (in addition to conventional box
diagrams) can
contribute significantly to the neutral kaon mass difference.
Unlike a superweak model \cite{W},
our description  of neutral kaons yields a non-zero value of
$\vert\epsilon^\prime/\epsilon\vert$ 
bounded above by $2\times10^{-3}$ \cite{FN}. 

Our model exhibits no strong $CP$ problem in tree approximation. The
determinants of ${\cal M}_{u,d}$ as
given by (\ref{Mass}) are real, so that  $\bar{\theta}=0$ at this order.
An upper limit on $x^2$ is
obtained from an examination of one-loop corrections to ${\cal M}_{u,d}$.
For each of these matrices, we have
\begin{equation}
\delta {\cal M} = \left( \begin{array}{cc} \delta
m & \delta F \cr
\delta G & \delta M \cr \end{array} \right),
\label{delta}
\end{equation}
from which we find:
\begin{equation}
\begin{array}{ccc}
\bar{\theta} & = & {\rm Arg\,Det}\left({\cal M} + \delta{\cal M}\right)\cr
 & = & {\rm  Im\,Tr}\left(\ln{{\cal M}(1+{\cal M}^{-1}\,\delta{\cal
M})}\right) \cr
 & \simeq &
{\rm Im\,Tr\,} {\cal M}^{-1}\,\delta{\cal M}. 
\end{array} 
\label{deth}
\end{equation}
The last equality in 
(\ref{deth}) is perturbative 
in the one-loop radiative 
corrections.
From (\ref{delta}) and (\ref{deth}) we deduce a useful expression 
for the one-loop correction to either of the phases $\bar{\theta}_{u,d}$:
\begin{equation}
\bar{\theta}\simeq {\rm Im\, Tr} \left(m^{-1}\delta m+ 
m^{-1}FM^{-1}\delta G+ M^{-1}\delta M\right)
\label{deltatheta}
\end{equation} 
An explicit examination of the several
bracketed terms in (\ref{deltatheta}) reveals
that all are  real except the first. The term $m^{-1}\delta m$ yields a
one-loop contribution to $\bar{\theta}=\bar{\theta}_u+
\bar{\theta}_d$ of order $\lambda
x^2/16\pi^2$, where 
$\lambda$ is the real coefficient of the Higgs interaction $\lambda
|\phi|^2|\chi|^2$. 
From $M_{\phi}^2 > (60\; {\rm GeV})^2$ and the 
constraint 
$\vert\langle\chi\rangle\vert < 2$~TeV \cite{FN}, we 
find $\lambda<10^{-3}$. With $x^2$ at its lower limit, $\lambda$ must be
thirty times smaller than this upper bound
to avoid a strong $CP$ problem.
 It would be unnatural to set $\lambda<10^{-5}$.
Thus, we find as a plausible range for $x^2$:
\begin{equation}
10^{-3}> x^2 > 3\times 10^{-5}.
\end{equation}

In an Aspon model, the quark mixing matrix is a complex unitary $4\times 4$
matrix $C_{\mu\nu}$.  Mixings of the conventional six quarks  with one
another are specified by the $3\times3$ matrix $C_{i,j}$, whose indices run
from 1 to 3. $C_{ij}$ is neither real nor unitary because of $\chi$-induced
mixings to the  undiscovered quark doublet. However, these terms are
$\sim\!x^2$. Thus $C_{ij}$, to a precision of at least 0.1\%, is a real
orthogonal matrix. It is a generalized Cabibbo matrix rather than
a Kobayashi-Maskawa matrix. This is unfortunate for the
search for $CP$ violation in the beauty sector, but has other observable
consequences. 

In the standard model, the Kobayashi-Maskawa matrix $V$ is complex and unitary.
The sides of the ``unitarity triangle'' are unity and:
\begin{equation}
R_b= \left\vert {V_{ud}V_{ub} \over V_{cd}V_{cb}} \right\vert,\qquad
R_t= \left\vert {V_{td}V_{tb} \over V_{cd}V_{cb}} \right\vert.
\label{tri}
\end{equation}
The value of $R_b$ has been measured. According to a recent review by M. Neubert
\cite{N}, 
$R_b= 0.35\pm 0.09$. The value of $R_t$ cannot be extracted from experimental
data alone. Appeal must
be made to a theoretical evaluation of the neutral $B$-meson mass
difference using the standard model. Neubert's analysis
yields $R_t= 0.99 \pm 0.22$. These results suggest a rather large value of
the $CP$-violating angle $\beta$, namely $0.34\le \sin{\beta}\le 0.75$.

In Aspon models, the matrix $C_{ij}$ is orthogonal up to terms 
of order $x^2$ arising from
mixings with unobserved quarks. Thus, we
anticipate no readily observable manifestations of $CP$ violation
in the beauty sector. Furthermore, the unitarity triangle must degenerate
into a straight line: $\vert R_b\pm R_t\vert =1$. In this case, we cannot
appeal to a theoretical calculation of the neutral $B$-meson mass
difference because it depends on unknown parameters. On the other hand,
the matrix $C_{ij}$ with the neglect of terms $\sim\!x^2$ involves only
three parameters. From data in hand, in this context, we obtain
$R_t=1-\rho$ in the Wolfenstein parametrization \cite{W1}, whence
$R_t=0.637\pm 0.09$. We note that present data yields $ R_b+R_t=0.99\pm 0.13$.
This result is compatible with an approximately
orthogonal mixing matrix  and hence with our model.

This work was supported in part by the U. S. Department of Energy under
Grant No. DE-FG05-85ER-40219, Task B, and by the National
Science Foundation under Grant No. PHY-92-18167.

\end{document}